\documentclass[twocolumn,showpacs,preprintnumbers,amsmath,amssymb]{revtex4-1}
\usepackage{amscd, graphicx}
\usepackage[all]{xy}

\makeatletter
\@addtoreset{equation}{section}
\makeatother
\usepackage{enumerate}

\def\cK{{\mathcal{K}}}
\def\cN{{\mathcal{N}}}
\def\cE{{\mathcal{E}}}
\def\cG{{\mathcal{G}}}
\def\pWS{{p_{\mathrm {WS}}}}
\def\EG{{E_{\mathrm {G}}}}
\def\kB{{k_{\mathrm {B}}}}
\def\erfc{{\mathrm {erfc}}}
\def\sigmaT{{\sigma_{\mathrm {total}}}}

\def\rhoHC{{\rho_{\mathrm {HC}}}}
\def\THT{{T_{\mathrm {HT}}}}
\def\THTc{{T_{\mathrm {HTc}}}}

\def\book#1{\rm{#1}, }
\def\paper#1{\textit{#1}, }
\def\jour#1{\rm{#1}, }
\def\yr#1{({\rm{#1}) }}
\def\vol#1{\textbf{#1}}
\def\pages#1{\rm{#1}}
\def\page#1{\rm{#1}}

\def\publaddr#1{\rm{#1}, }
\def\publ#1{\rm{#1}, }
\def\by#1{{\rm{#1}, }}

\begin{document}


\title{
A novel conductivity mechanism of highly disordered
carbon systems based on an investigation of graph zeta function}

\author{Shigeki Matsutani${}^{1)}$ and Iwao Sato${}^{2)}$}
 \affiliation{
1) National Institute of Technology, 
Sasebo College, 1-1 Okishin, Sasebo, 857-1193, Japan}
\affiliation{
IMI, Kyushu University, Motooka 744, Nishi-ku, Fukuoka 819-0395, Japan}
 \affiliation{
2) National Institute of Technology, 
Oyama College, Oyama 323-0806, Japan}
 \email{smatsu@sasebo.ac.jp, isato@oyama-ct.ac.jp}

\date{\today}

\begin{abstract}
In the previous report [Phys. Rev. B {\bf{62}} 13812 (2000)],
by proposing the mechanism
under which electric conductivity
is caused by the activational hopping conduction with
the Wigner surmise of the level statistics,
the temperature-dependent of electronic conductivity of a 
highly disordered carbon system 
was evaluated including apparent metal-insulator transition.
Since the system consists of small pieces of graphite,
it was assumed that the reason why the level statistics appears 
is due to the behavior of the quantum chaos in
each granular graphite.  
In this article, we revise the assumption and show another origin of the Wigner
surmise, which
is more natural for the carbon system based on a recent investigation
of graph zeta function in graph theory.
\end{abstract}

\maketitle

\section{Introduction}
In the previous reports \cite{MS1,MS2},
the temperature-dependence of the 
electric conductivity (TDEC)
of activated carbon fibers (ACFs) was investigated,
 which are
known as highly disordered systems (HDSs) \cite{DDSSG}. 
A mechanism of activational hopping conductivity 
for the disordered systems was proposed
that thermal excitations with the Wigner surmise, 
\begin{gather}
\pWS(\EG) d \EG =
2(\EG/\alpha^2) \cdot
\exp(-\EG^2/\alpha^2) d\EG,
\label{eq:WS}
\end{gather}
contribute to the conductivity. 
The mechanism provides the TDEC which is in good agreement with 
the experimental results of Kuriyama \cite{K} and
reproduces the (apparent) metal-insulator transition
in \cite{KD}.

Further in \cite{MS1,MS2},
it was microscopically assumed that
the Wigner surmise comes from the quantum chaos in the
small particles because the ACFs consists of small pieces of 
graphite \cite{DDSSG}; each piece of small graphite might be regarded as 
a quantum box and it is known that 
its energy eigenvalue is described well by the random matrix theory (RMT) 
and obeys the Wigner surmise 
(\ref{eq:WS})
 \cite{M}.

In this article, we revise the assumption and 
employ another origin of the Wigner surmise
based on the recent results of the graph theory
on Ihara's zeta function \cite{T}.
The new origin is much more natural for the
ACFs.

\bigskip
The electrical and structural properties of the ACFs were studied
by Kuriyama and Dresselhaus \cite{K,KD} and others \cite{DDSSG,DFRVKDE}.
Since the ACF consists of small pieces of
graphite, the X-ray diffraction and Raman spectra show that
the heat-treated process (HTP) of the ACFs modifies the structure
and the size of the pieces drastically changes \cite{DDSSG,DFRVKDE}.
Kuriyama studied the TDEC of the ACFs and its dependence on the
HTP experimentally \cite{K}, and 
found the fact that if the density of states (DOS) of the
activational energy is given by a $\Lambda$-shape and the conductivity
is proportional to the thermal activation,
the estimation reproduces his own experimental results of the 
TDEC.

It implies that
the previous reports \cite{MS1,MS2} gave
a microscopic picture of Kuriyama's 
mechanism based on the quantum chaos as
mentioned above.

It is well-known that the RMT coming from the quantum chaos also
represents the behaviors  of the Riemann zeta function
in the number theory \cite{T}.

Recently Newland showed numerically that
even Ihara's zeta function in the graph theory also obeys
the Wigner surmise \cite{T}. Since the zeta
function is determined
by the spectrum of the adjacency matrix (SAM) in graph theory,
it was also found numerically that the 
SAM of a graph is also governed by the Wigner surmise.
These results were obtained from the motivation of pure mathematics.
However from a physical point of view, 
it means that the energy gaps  of the tight binding model
of certain materials with chemical bonding
obey the Wigner surmise  (\ref{eq:WS}).
Thus we employ this picture in this article.


\section{Results of the Previous report}

Here we will review the formula in \cite{MS1,MS2}
on the TDEC in HDCs or the ACFs 
with the temperature $\THT$ of the HTP as a parameter.

\begin{figure}[h]
 \begin{center}
\includegraphics[width=5cm]{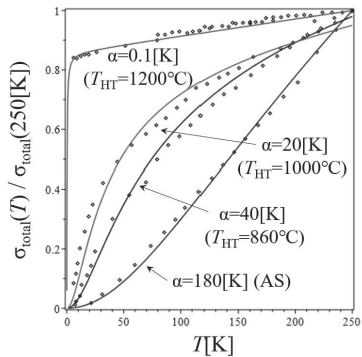}
 \newline
 \end{center}
\vskip -0.5cm
\caption{
TDEC: 
The dotted points show the 
experimental results of TDEC $\sigmaT(T)/\sigmaT(250[K])$
for as-prepared (AS) and heat-treated ACFs 
and curves following 
(\ref{eq:sigma1}) 
for fitting parameters
$\alpha$ and $\delta^{-1} =  
1.5\times 10^3[K^{-1}]$.}
\label{fig:PResult}
\end{figure}

As mentioned in \S I,
by employing the Wigner surmise (\ref{eq:WS}) as the DOS,
the total conductivity is expressed by
\begin{equation}
\begin{split}
\sigmaT (T) &= 
\sigma_0(T)\int d \EG\  \pWS(\EG) \exp(-\EG/\kB T)\\
&=\sigma_0(T) 
\left[1-\sqrt{\pi}
\frac{\alpha}{T}\exp\left(\frac{\alpha^2}{T^2}\right)
\cdot \erfc\left(\frac{\alpha}{2T}\right)
\right],\\
\end{split}
\label{eq:sigma1}
\end{equation}
where $\alpha$ is a fitting parameter depending on $\THT$,
and  
$\sigma_0(T)$  weakly depends on the temperature $T$,
$\sigma_0(T)=\sigma_0^{(0)} 
(1+\delta T)$.
Then (\ref{eq:sigma1}) 
reproduces well the experimental results \cite{K}
 including the insulator-metal transition \cite{KD} as in 
FIG.~\ref{fig:PResult}.

As mentioned in \S I, it was considered in \cite{MS1,MS2}
that the 
origin of the Wigner surmise 
(\ref{eq:WS})
was quantum chaos for
the quantum boxes since
the graphite pieces of the ACF might be regarded as
small granular particles.
For example, the quantum system of stadium
is governed by the RMT.
The Wigner surmise naturally appears in the RMT
since each level is repulsive
in \cite{B,M}.

\section{Electric structure of graphite}

The electronic property of graphite has been studied well.
The tight-binding approximation (TBA) of 
$\pi$-electron distribution of graphite as a 2D
honeycomb lattice was studied in \cite{W,CR}.
Coulson and Taylor also considered effects of overlap integrals
and 3D effect \cite{CT}.
Zunger investigated a more realistic model \cite{Z}.

In these studies, the infinite size of lattice
or pure crystal of graphite was assumed.
On the other hand, 
as in \cite[p.71]{DDSSG} and \cite{DFRVKDE},
the ACFs consists of small pieces of graphite 
and the HTP of the ACFs modifies the structure
and the size of the pieces drastically.

The boundary of the graphite piece is not stable due to dangling bonds.  
Since the phonon is easily excited there,
the coherency of the electron wave function 
is loosed there.
Since the ACF is considered 
as a collection of small graphite pieces, the electronic
band structure is given as that of independent pieces
as in \cite[p.153-161]{DDSSG}.

In the picture, since the shape of the piece is crucial,
we go back to the TBA as a simple approximation \cite{SL}.

\section{The TBA and theory of graph}

\begin{figure}[h]
 \begin{center}
\includegraphics[width=6cm]{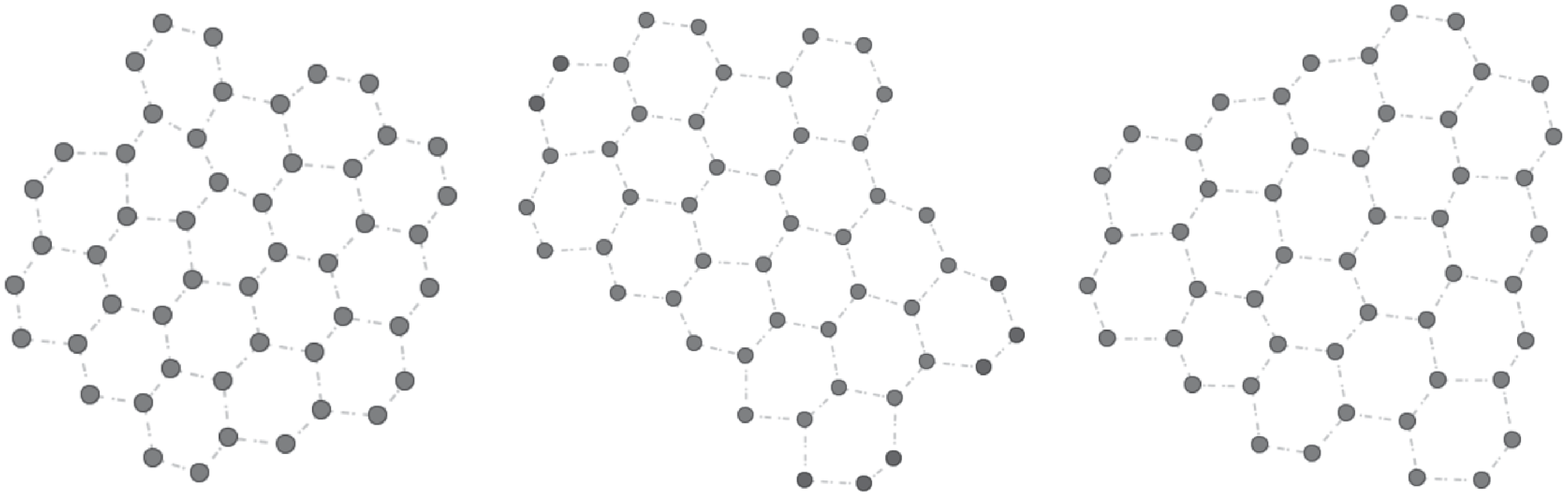}
 \newline
(A1):50\hskip 1.5 cm 
(A2):50\hskip 1.5 cm 
(A3):50
\newline
\includegraphics[width=6cm]{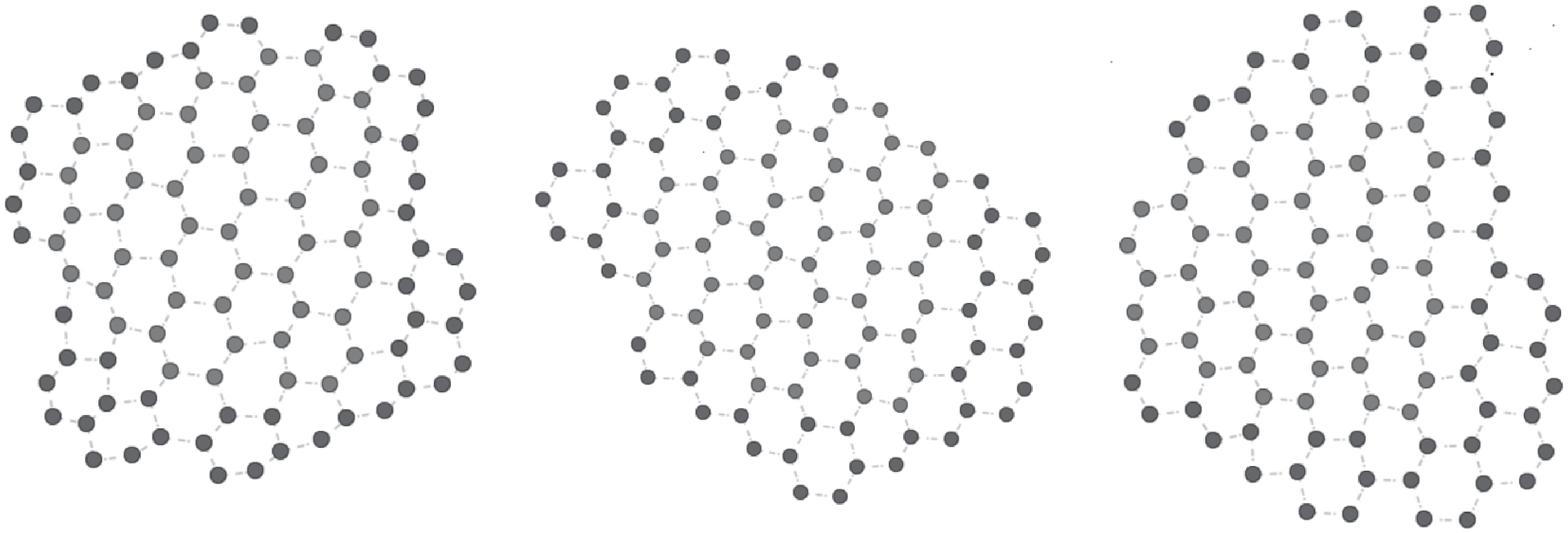}
 \newline
(B1):100\hskip 1.5 cm 
(B2):100\hskip 1.5 cm 
(B3):98
\newline
\includegraphics[width=6cm]{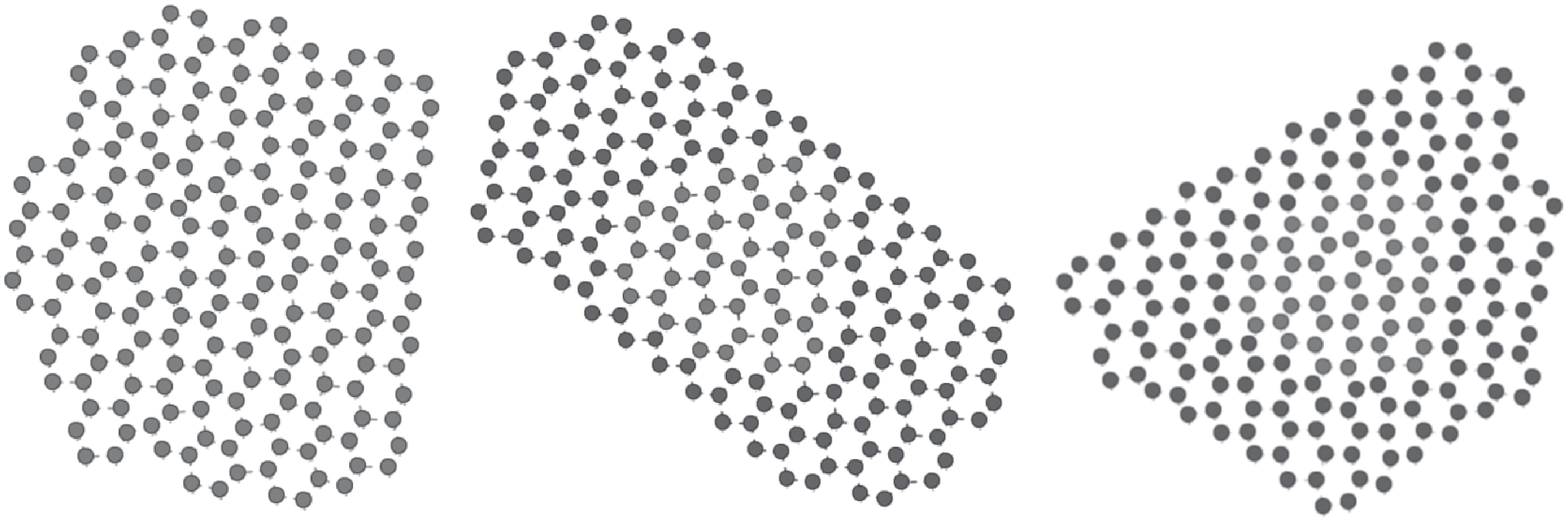}
\newline
(C1):199\hskip 1.5 cm 
(C2):202\hskip 1.5 cm 
(C3):201
\newline
\end{center}
\vskip -0.5cm
\caption{
The graphs $\cG$'s with numbers of nodes $\cN_\cG$.
}
\label{fig:GFs}
\end{figure}

We consider the electronic structure of
the small graphite pieces  in the framework of
the TBA.
For a given graph $\cG$, e.g. FIG.~\ref{fig:GFs},
let the set of nodes in $\cG$ denoted by $\cN_\cG$ and 
that of edges in $\cG$ by $\cE_\cG$.
For the real numbers $\varepsilon_0$ and $\gamma_0$
$= 3.16$[eV] \cite[p.160]{DDSSG},
the hamiltonian of $\cG$ is given by
$
 H = \varepsilon_0 D_\cG - \gamma_0 A_\cG,
$
where $D_\cG$ is the diagonal matrix and 
$A_\cG$ is the adjacency matrix of the graph $\cG$;
for every $i,j \in \cN_\cG$,
$(D_\cG)_{ij} = \delta_{ij}$, 
and 
$(A_\cG)_{ij} = 1$ if there is an edge $(i,j)$ in $\cE_\cG$
otherwise vanishes \cite{D,Cas}.
$H$, $D_\cG$, and $A_\cG$ are $N\times N$-matrices for
$N=|\cN_\cG|$.
The spectrum of $H$ is determined by the SAM $A_\cG$ \cite{D}.

In the graph theory, 
Ihara's zeta function and the SAM 
have been studied as a graph theoretic version of the 
Riemann hypothesis \cite{KSu,T}.
As in \cite{M}, 
since many observations show 
that mathematical properties of 
the Riemann zeta function associated with Riemann hypothesis
are expressed well by the RMT,
the graph zeta function and 
SAM should be also described
by the RMT.
It is shown that for a certain 
(random) graph, the Wigner semicircle law governs the 
SAM \cite{S,T}.

Further Newland, in his thesis 2005, showed by numerical
computations that in a certain graph,
the spectra of the adjacency matrices and
Ihara's zeta functions obey the Wigner surmise 
(\ref{eq:WS})
\cite[p.41]{T}.

\bigskip

\begin{figure}[h]
 \begin{center}
\includegraphics[width=5cm]{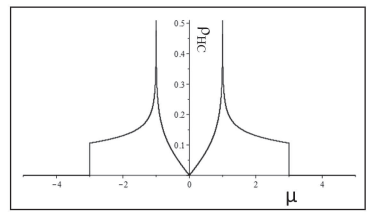}
 \newline
 \end{center}
\vskip -0.5cm
\caption{
The DOS of the 2D Honeycomb lattice}
\label{fig:LGF}
\end{figure}

The SAM of the infinite
2D graphite lattice or  the honeycomb lattice
was obtained by Coulson and Rushbrooke \cite{CR} and,
later, was precisely studied by Horiguchi in terms 
of the lattice Green function
method \cite{H,Cas}.
The DOS is explicitly written by
\begin{equation}
\rhoHC(\mu) = 
\left\{\begin{matrix}
\frac{2\sqrt{\mu}} 
{3\sqrt{3}\pi^2} \cK(\kappa(\mu))& \mbox{for}& 1 <|\mu|<3\\
\frac{2\sqrt{\mu}} 
{3\sqrt{3}\pi^2\kappa(\mu)} 
\cK\left(\frac{1}{\kappa(\mu)}\right)
& \mbox{for}& |\mu|\ge1\\
\end{matrix}\right.,
\label{eq:LGF}
\end{equation}
where $\cK$ is the Jacobi elliptic integral of the 
first kind and
${\kappa(\mu) = \frac{1}{4}\sqrt{
\frac{ (1+|\mu|)^3(3-|\mu|)}{|\mu|}}
}$, as in FIG.~\ref{fig:LGF}.
It is also noted that
Ihara's zeta function of the infinite lattice is related to 
the elliptic integral \cite{C}.

The SAM of a small piece of
graphite asymptotically approaches
to the DOS (\ref{eq:LGF}) when its size approaches to $\infty$.
It is contrast to the fact that
for a certain class of the random graph, the 
asymptotic behavior of the DOS obeys the semicircle
law \cite{S}. 

However it is expected that
the density of the level-spacing of
the SAM is determined by a function similar to the Wigner
surmise (\ref{eq:WS}) because 1) the degenerate states are naturally
avoided if there is no global symmetry and 2) the average of the 
level spacings must be determined by the insertions of 
$|\cN_\cG|$ points
into the region $[-3,3]$.

\section{Conductivity of the ACFs} \label{sec:NewT}

Let us employ the ansatz that the level statistics of
the graphite piece obeys the 
Wigner surmise (\ref{eq:WS}).
Though the gap of the first excited state from the Fermi level 
(the gap between
HOMO (highest occupied molecular orbital) and
LUMO (lowest unoccupied molecular orbital))
is concerned only, 
it is natural to assume that the gap also obeys the Wigner surmise
(\ref{eq:WS})
from recent development of the graph theory mentioned above \cite{T}.

In the framework of the TBA,
the occupied states depend on the number of the carbon atoms
and due to the spin effect; there is the gap if the number is even
whereas the state is not filled if the number is odd.
Though depending on the parity of each graphite piece,
there appears the following picture as a large resistance case.

\begin{figure}[h]
 \begin{center}
\includegraphics[width=6cm]{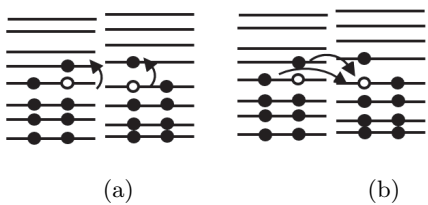}
 \newline
(a) \hskip 3cm
(b)
 \end{center}
\vskip -0.5cm
\caption{
Hopping process from (a) to (b).
}
\label{fig:HPab}
\end{figure}

Let us consider the hopping phenomena between
two adjacency graphite pieces.
As in illustrated  in FIG.~\ref{fig:HPab},
in order that an electron in a piece hops to its adjacency one,
the electron must jump to excited states as the first step.
Under the electric field, the possibility of hopping 
is influenced and we have finite conductivity.

Let us consider that the ACF is arranged between two electrodes.
From one electrode to another, there are possible electric paths
$\Gamma_0:=\{C_a\}$ \cite{MS,MS1,MS2}. 
Let $\sigma_{i,j}$ be the local 
conductivity of adjacency graphite pieces belonging to the path $C_a$.
The conductivity $\sigma_{C_a}$ along the path $C_a$
could be approximated by 
$$
\sigma_{C_a} = \left(\sum_{i} \frac{1}{\sigma_{i,i+1}}\right)^{-1}
\approx \left(\max_{i} \frac{1}{\sigma_{i,i+1}}\right)^{-1}.
$$ 
Then the total conductivity is simply obtained by 
the summation over all paths,
$
\sigmaT = \sum_{C_a \in \Gamma} \sigma_{C_a}.
$ It means  (\ref{eq:sigma1}) and
the above picture shows its microscopic origin.

\section{Numerical Study}

%

\begin{figure}[h]
 \begin{center}
\includegraphics[width=8.5cm]{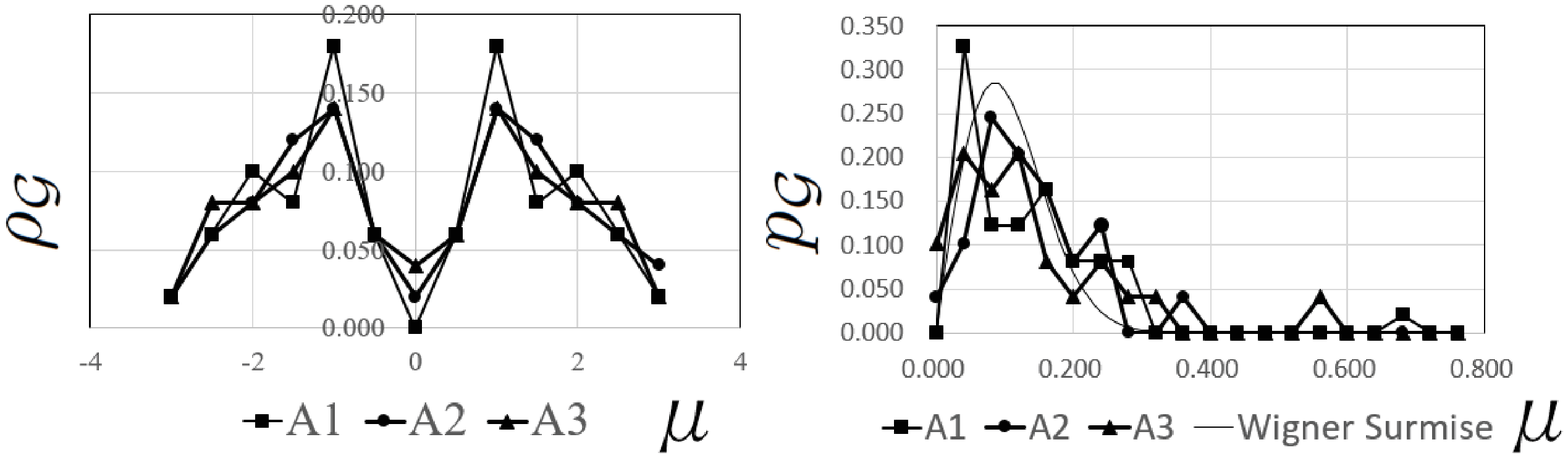}
 \newline
\hskip 3 cm 
(a)\hskip 3 cm 
(b)
\newline
\includegraphics[width=8.5cm]{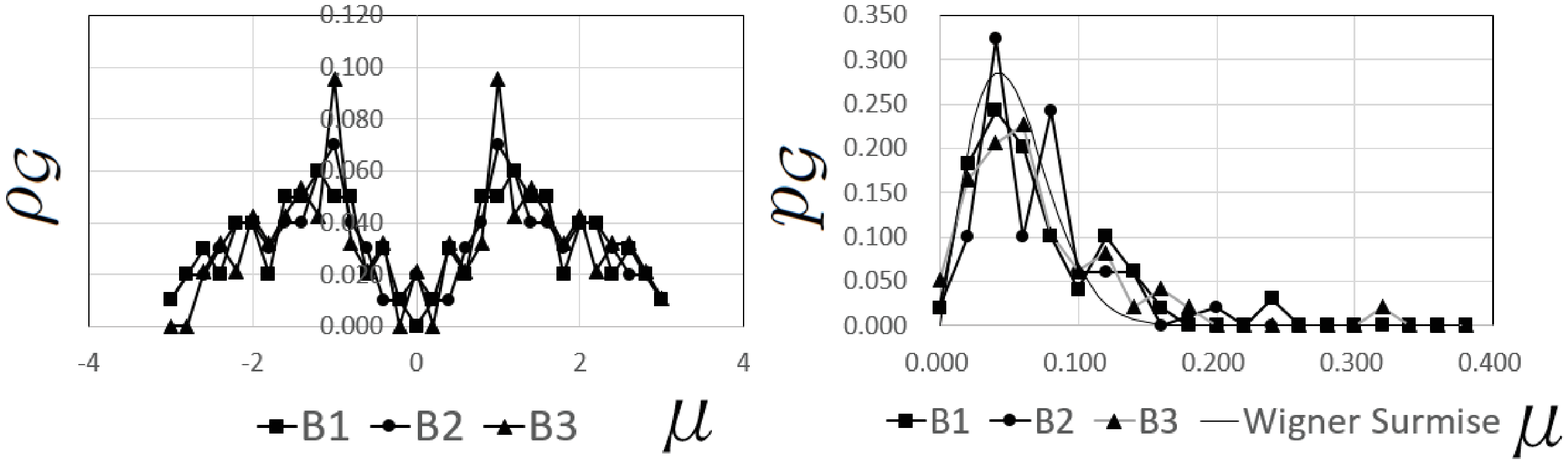}
 \newline
\hskip 3 cm 
(c)\hskip 3 cm 
(d)
\newline
\includegraphics[width=8.5cm]{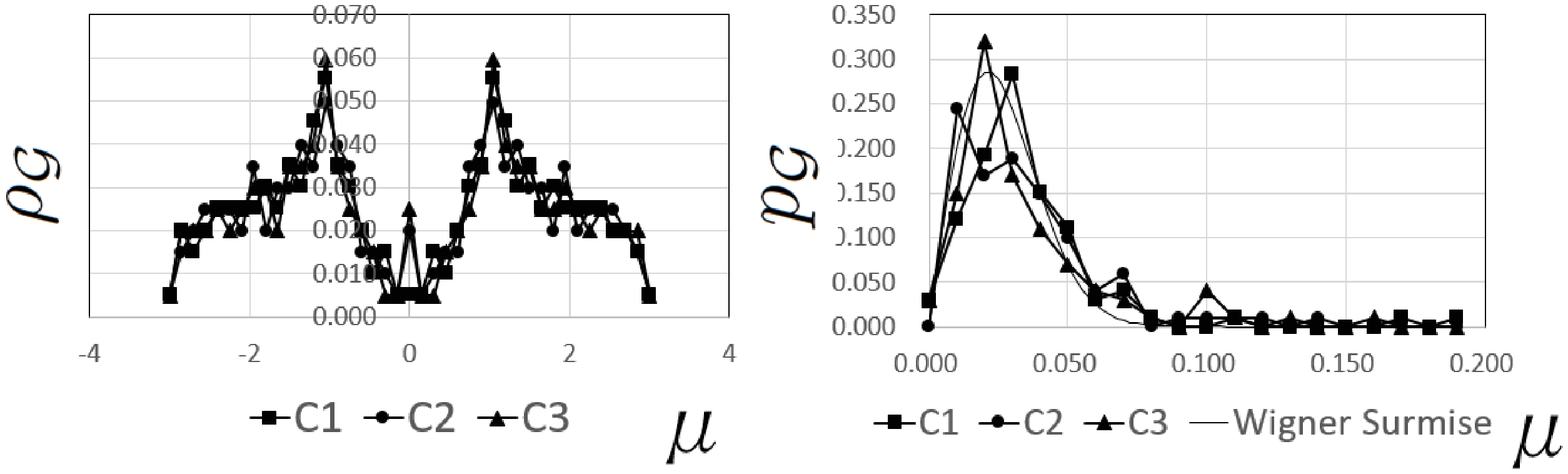}
 \newline
\hskip 3 cm 
(e)\hskip 3 cm 
(f)
 \end{center}
\vskip -0.5cm
\caption{
Numerical Results of $\rho_\cG$
and $p_\cG$.
}
\label{fig:Hp}
\end{figure}

We, first, numerically showed the justification of the ansatz
in \S V for graphs given as FIG.~\ref{fig:GFs}
using the software {\it{Graphtea}};
we computed their SAM $\{\mu\}$ by 
solving $\mathrm{det}(\mu I - A_\cG)=0$ numerically.
Let the distribution of the SAM denoted by $\rho_\cG$
and that of level spacings by $p_\cG$. 
The results are illustrated in 
FIG.~\ref{fig:Hp}, which shows that
$\rho_\cG$ approach to $\rhoHC$ (\ref{eq:LGF}) asymptotically 
and $p_\cG$ is similar to $\pWS$ due to the level repulsion.
In other words, our ansatz in \S V is natural.
In fact, FIG.~\ref{fig:Hp}  exhibits that
$p_\cG$'s are approximated by the Wigner surmise
(\ref{eq:WS}).

We computed the center of the gravity for 
$p_\cG$'s, i.e.,
${\int \mu p_\cG(\mu) d\mu}$,
and the effective $\alpha$ for $\gamma_0 = 3.16$[eV]
noting 
${
\int_0^\infty \pWS(\mu) \mu d\mu = \frac{\sqrt{\pi}}{2} \alpha\
}$
as in Table 1: 
\begin{table}[htb]
Table 1. Center of gravity of $p_\cG$ and $\alpha$ [K]\\
  \begin{tabular}{|c|r|r|r|r|r|} \hline
          &  1  &  2 &  3  & average & $\alpha$[K]\\
  \hline
  \hline
   A-type & 0.136 &0.135 &0.134& 0.135 & 5586\\
  \hline
   B-type & 0.069 &0.068 &0.069& 0.069 & 2842\\
  \hline
   C-type & 0.035 &0.035 &0.034& 0.035 & 1436\\
  \hline
  \end{tabular}
\end{table}

Using the least mean square method,
the relation between 
$\alpha$ and the number $N=|\cN_\cG|$
is evaluated as $\alpha = 2.8 \times 10^5 / N$[K]
from Table 1.
With the estimation $N = \pi (R/a)^2$ for the
radius $R$ and $a = 0.142$[nm], 
the fitting parameters $\alpha$'s in FIG.~\ref{fig:PResult}
are reduced to the effective radii $R$'s as in Table 2.

\begin{table}[htb]
Table 2. $\alpha$ and $R$ of  the ACF\\
  \begin{tabular}{|c|c|r|r|r|} \hline
   &       &  $\alpha$  [K] & N&$R$[nm]\\
  \hline
case I &   AS &  180 & $1.6\times 10^3$ &3.2\\
  \hline
case II &   $\THT=860$[${}^\circ$C] & 40 & $7.0\times 10^3$ &6.7\\
  \hline
case III &   $\THT=1000$[${}^\circ$C] & 20 & $1.4\times 10^4$ &9.5\\
  \hline
case IV &   $\THT=1200$[${}^\circ$C] & 0.1 &$2.8\times 10^6$ & 1.3 $\times 10^2$\\
  \hline
  \end{tabular}
\end{table}


Since X-ray analysis shows that there is 1.1[nm] peak which
corresponds to the length along $c$-axis of three stratified
graphite sheets \cite{DDSSG,DFRVKDE}.
Thus by considering 3D effect,
$R$'s of case I-III in Table 2 should be divided by several numbers.
From p.71 in \cite{DDSSG},
the structure of the ACF strongly depends on $\THT$ of HTP and
 especially around $\THTc=1500$ [${}^\circ$C], the structure drastically
changes and has 3D property 
as a kind of structural phase transition.
However the critical temperature $\THTc$ also depends on
the material origin of the ACFs. Thus the ACF of case IV 
may have the 
3D structure
and if we use $N= 4\pi R^3/3a^3$, $R= 12[nm]$.
Then these estimations of $R$'s are compatible with TEM data in
\cite{DDSSG} and it also turns out that the (apparent) metal-insulator 
transition 
might come from the structural transition due to the HTP.

\section{Discussions}
In this article, we show that the SAM
reproduces the conductivity of the disordered carbon system.
We conclude that our picture is natural to the TDEC of the ACFs.
In the series of works \cite{IKMM,M,MO,MS},
One of the authors (S.M.) has been showing that some of physical phenomena are
expressed well by pure mathematical results.
Since the SAM appears in theory of the
 graph zeta function
which has been studied as a graph theoretic version of
 the Riemann hypothesis \cite{KSu,S,T,VN}, the
TDC of the ACFs is one of such cases.

However the studies on the Wigner surmise
of SAM are not sufficient.
Especially the gap between HOMO and LUMO of 
pieces of graphite should be studied more systematically \cite{M2}.
Further since there are studies of more realistic computation of
electronic structure of graphite pieces \cite{Ca,Cas},
it is expected that the statistical property of gaps for these systems is
evaluated in future. 

Since recently it is found that 
Ihara's zeta functions naturally appears in 
quantum walks \cite{KSa}, it means that 
this investigation might show another possibility of
quantum walk.

\bigskip
\noindent
{\bf{Acknowledgment:}}
We are grateful to Professor Norio Konno 
for helpful discussions.
We also acknowledge the graph software {\it{Graphtea}} developed in
Graphlab in Sharif University of Technology.
This work
was supported by the Grant-in-Aid for Scientific Research (C) 
of Japan Society for the Promotion of Science (Grant No. 16K05187 (S.M.) 
and Grant No. 15K04985 (I.S.)).

\bigskip


\end{document}